\documentclass[a4paper,12pt]{article}

\usepackage[dvipsnames]{xcolor}
\usepackage{amsmath}
\usepackage{amssymb}
\usepackage{dsfont}
\usepackage[english]{babel}
\usepackage[utf8]{inputenc}
\usepackage{times}
\usepackage{graphics}
\usepackage{graphicx}
\usepackage[T1]{fontenc}
\usepackage[official,right]{eurosym}
\usepackage{url}
\usepackage{eso-pic}
\usepackage{tikz}
\usetikzlibrary {arrows.meta,bending,positioning}
\usetikzlibrary{shapes.geometric, arrows}
\usepackage{eurosym}
\usepackage{enumerate,rotating,multirow,xcolor,booktabs,placeins}
\usepackage{algorithm}
\usepackage{algpseudocode}
\usepackage{appendix}
\usepackage{fancyhdr}
\usepackage{natbib}

\def\qed{\hbox to\hsize{\hfill\vrule height 1.6ex width 1.5ex depth -.1ex}}

\usepackage{enumitem}
\newlist{steps}{enumerate}{1}
\setlist[steps, 1]{label = Step \arabic*:}

\title{Integrating problem structuring methods with formal design theory:  collective \\ water management policy design in Tunisia}
\author{H. Berkay Tosunlu$^\dag$, Joseph H.A. Guillaume$\ddag$, \\ Alexis Tsouki\`as$^\dag$, Emeline Hassenforder$+$, Samia Chrii$*$, \\ Houssem Braiki, Irene Pluchinotta$**$ \\ \small{$^\dag$ CNRS-LAMSADE, PSL, Universit\'e  Paris Dauphine} \\ \small{$\ddag$ Fenner School of Environment and Society, Australian National University} \\ \small{$+$ CIRAD, UMR G-EAU, Tunis} \\ \small{$*$ Cirad, INRGREF} \\ \small{$**$ Institue for Environmental Design and engineering, University College London}}
\date{}

\begin{document}

\thispagestyle{empty}

\enlargethispage*{8cm}
 \vspace*{-38mm}

\AddToShipoutPictureBG*{\includegraphics[width=\paperwidth,height=\paperheight]{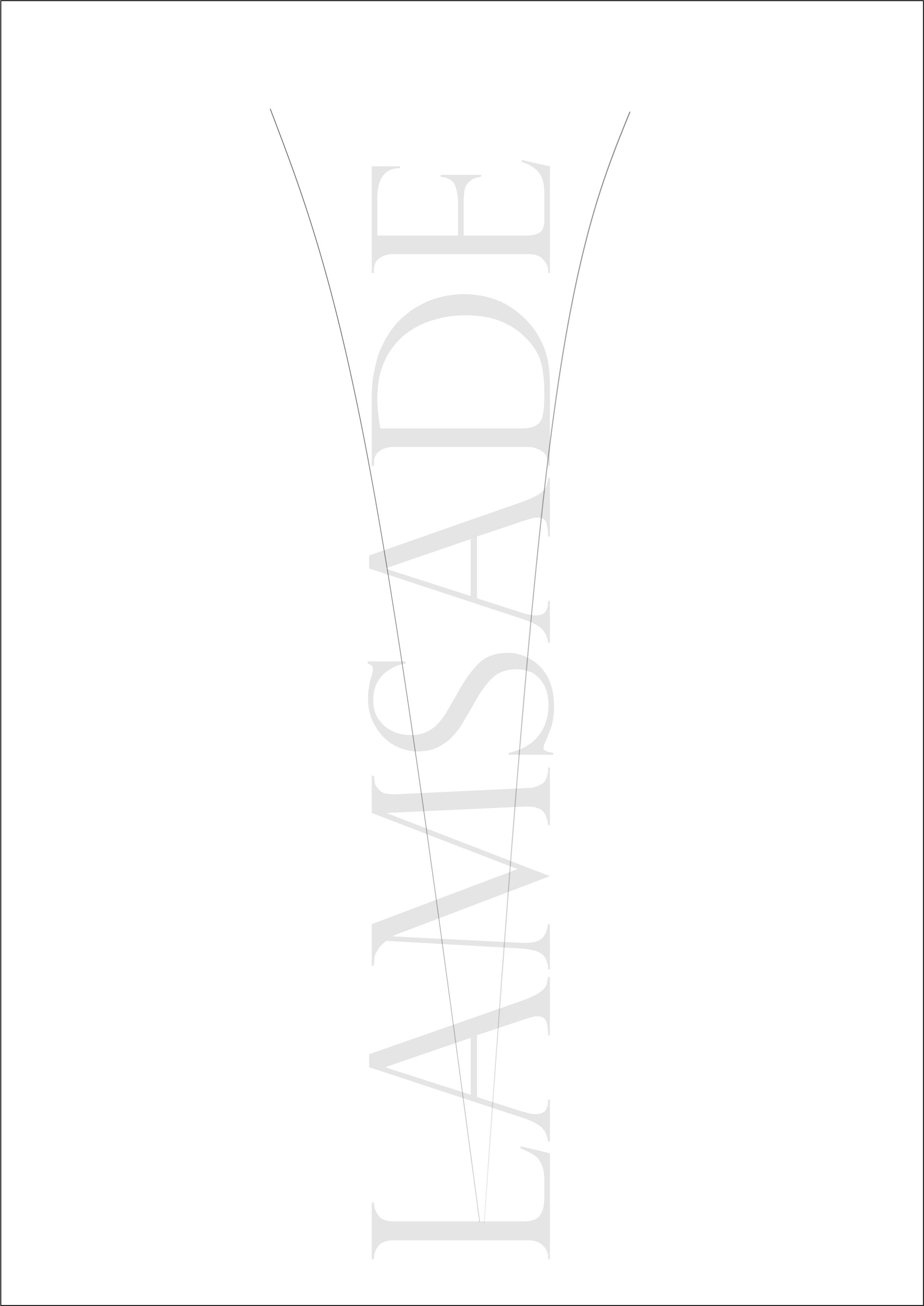}}

\begin{minipage}{24cm}
 \hspace*{-28mm}
\begin{picture}(500,700)\thicklines
 \put(60,670){\makebox(0,0){\scalebox{0.7}{\includegraphics{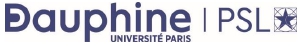}}}}
 \put(60,70){\makebox(0,0){\scalebox{0.3}{\includegraphics{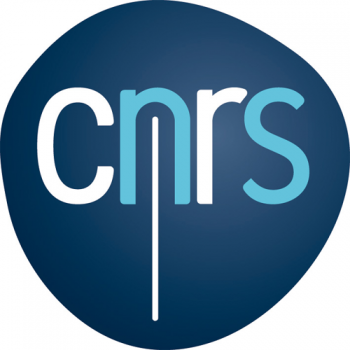}}}}
 \put(320,350){\makebox(0,0){\Huge{CAHIER DU \textcolor{BurntOrange}{LAMSADE}}}}
 \put(140,10){\textcolor{BurntOrange}{\line(0,1){680}}}
 \put(190,330){\line(1,0){263}}
 \put(320,310){\makebox(0,0){\Huge{\emph{410}}}}
 \put(320,290){\makebox(0,0){September 2024}}
 \put(320,230){\makebox(0,0){\Large{Integrating problem structuring methods with}}}
 \put(320,210){\makebox(0,0){\Large{formal design theory: collective water management}}}
 \put(320,190){\makebox(0,0){\Large{policy design in Tunisia}}}
 \put(320,100){\makebox(0,0){\Large{H. Berkay Tosunlu, Joseph H.A. Guillaume }}}
 \put(320,80){\makebox(0,0){\Large{Alexis Tsouki\`as, Emeline Hassenforder, Samia Chrii}}}
 \put(320,60){\makebox(0,0){\Large{Houssem Braiki, Irene Pluchinotta}}}
 \put(320,670){\makebox(0,0){\Large{\emph{Laboratoire d'Analyse et Mod\'elisation}}}}
 \put(320,650){\makebox(0,0){\Large{\emph{de Syst\`emes pour l'Aide \`a la D\'ecision}}}}
 \put(320,630){\makebox(0,0){\Large{\emph{UMR 7243}}}}
\end{picture}
\end{minipage}

\newpage

\addtocounter{page}{-1}

\maketitle

\abstract{Groundwater management, especially in regions like Tunisia, is challenging due to diverse stakeholder interests and the dry structure of climate, which is extremely challenging for the sustainability of water resources. This paper proposes an innovative approach to policy design by merging Problem Structuring Methods (PSMs) and the Policy-Knowledge, Concepts, Proposals (P-KCP) methodology. Utilizing cognitive maps and value trees, the study aims to generate new collective groundwater management practices. Bridging decision theory and design theory, the study addresses the gap in new alternative generation and highlights the P-KCP's role in innovative policy design. Integrating PSMs and C-K theory, the framework expands policy alternatives and advocates for participatory approaches. It emphasizes adaptability across contexts, provides replicable process descriptions, and encourages the creation of unconventional policy solutions. Ultimately, this comprehensive framework offers a practical guide for policy innovation and collaboration.}

\newpage

\section{Introduction}\label{intro}

\footnote{Disclosure of interest: There are no relevant financial or non-financial competing interests to report. No funding was received}
The aim of this paper is to present innovative tools for public policy design combining problem structuring methods (essentially cognitive maps and value trees) with formal design theory. Our proposal consists in transforming cognitive maps (CM) \citep[see][]{eden1988cognitive,eden2004cognitive} into value trees (VT) \citep[see][]{von1987value,keeney1994creativity,poyhonen1998notes} and integrating them with formal design theory, especially Concept-Knowledge (C-K) theory \citep[see][]{hatchuel2003new,hatchuel2009ck}. This approach will help in generating effective policy solutions by combining descriptive stakeholder perspectives with a structured, design-oriented methodology. Additionally, we aim to support the  Policy-Knowledge, Concepts, Proposals (P-KCP) methodology by offering CM and VT as foundational tools. Since creating a concept tree is complex and time-consuming, we propose using the VT as a starting point for the P-KCP process.

This study aims to present a comprehensive framework that addresses three crucial aspects of policy design: adaptation to different contexts, facilitation of replicability, and an innovative approach to policy design. Firstly, we carefully tailored the policy design tool to suit various contexts, taking into account the unique challenges and characteristics of each setting. This adaptability ensures the methodology remains relevant and effective in different policy domains. Secondly, to promote replicability, the study provides a detailed description of each phase involved in the policy design process. From problem formulation and stakeholder engagement to the co-evolution of concept and knowledge spaces, each step is outlined in a clear and systematic manner, allowing other researchers and practitioners to easily replicate and customize the approach for their own contexts. Lastly, the framework emphasizes the importance of innovative policy design by integrating PSMs and the P-KCP methodology. By doing so, the study encourages the generation of new and out-of-box policy alternatives, moving beyond the evaluation of known options. This participatory approach fosters collaboration and empowers stakeholders to explore previously unknown policy solutions. In conclusion, this comprehensive framework not only addresses the complexities of diverse contexts but also offers a practical and adaptable guide for driving innovation in policy-making.

Our framework has been tested in an action research setting concerning groundwater management in the Limaoua region, near Sfax in Tunisia. Managing groundwater resources in Tunisia is a critical task due to the diverse interests of involved stakeholders and the fact that water demand is higher than water availability. The arid conditions, coupled with a heavy groundwater reliance on agriculture, underscore the urgent need for effective groundwater management strategies. These strategies must balance the ecological health and agricultural productivity of the region, which is increasingly threatened by overexploitation and climatic uncertainties. The complexities of managing such a vital resource are magnified by the varying needs and perspectives of local farmers, policymakers, and other stakeholders, highlighting the necessity of a sustainable approach to water use. Farmers need to water their crops and, therefore, want to use the water resources to satisfy the needs of their crops. However, water authorities need to preserve the water resource. In this kind of conflicting interest, there is a need for policy design focused on sustainability.

In most places of Tunisia, agricultural intensification over the years has led to environmental degradation, including significant depletion of water tables and the deterioration of water quality. These challenges, further intensified by the differing priorities of local farmers, policymakers, and environmental advocates, highlight the urgent need for a collaborative approach to achieve groundwater sustainability. Drawing inspiration from a previous collective management model \citep[see][]{frija2016gestion}, this study investigates participatory models tailored to Limaoua's unique context. Our goal is to develop sustainable and equitable management solutions that address the region's pressing groundwater issues effectively. For this purpose, we present an experiment consisting of using innovative policy design tools for participatory natural resource management.

\section{Background}\label{bkg}

\subsection{PSMs: Cognitive maps and value trees}\label{psms}


Soft Operational Research (OR) methods, also known as Problem Structuring Methods (PSMs), are invaluable tools in addressing the multifaceted nature of decision-making in complex scenarios  \citep[see][]{rosenhead1996s,Ackoff79a,Ackoff79b}. PSMs embrace the ambiguity and subjective elements inherent in real-world situations involving multiple stakeholders with diverse interests \citep[see][]{mingers2004problem}.

PSMs primarily serve as descriptive tools, detailing who faces which problems and why, thus helping stakeholders develop comprehensive assessments and establish common ground \citep[see][]{ackermann2012problem}. These methods would benefit from incorporating a more design-oriented approach; they do not necessarily propose practical solutions or pathways to resolve the identified problems, particularly those requiring 'out-of-the-box' thinking or addressing 'wicked situations' \citep{simon1969sciences}.

Exceptions like the ``Strategic Choice Approach'' \citep{FriendHickling87} and ``Value Focused Thinking'' \citep{Keeney92} exist, but they also struggle with understanding the structural dynamics of problems and decision aiding complexities \citep[see][]{Tsoukias07aor,Tsoukias08ejor}. Recent research \citep{ColorniTsoukias2020,ferretti2019studying,pluchinotta2019design} calls for a more 'design-oriented' approach in decision support, emphasizing the need for methods that not only describe problems but also guide the generation of alternatives.

In a recent paper \citep{tosunlu2023conflict} we suggest a new approach (and a related algorithm) allowing to transform a cognitive map to a value tree the latter being considered as the driver of decision making behaviour.

In our examination of problem-solving methodologies, we utilize Kelly's theory of humans as innate problem solvers \citep[see][]{kelly1955psychology}, which contributes to our comprehension of cognitive mapping as outlined by \cite{eden1988cognitive}. Eden’s approach emphasizes the subjective nature of problem handling, focusing on the problem owner's personal understanding, values, beliefs, and objectives \citep[see][]{eden1994cognitive}. While cognitive mapping captures the perspective of the problem owner, it does not inherently provide direct strategies for complex situations, requiring additional steps in the problem-solving process. However, as highlighted by \citep{graf2007counselling}, creativity is crucial for conflict transformation.

Conversely, value trees offer a structured approach rooted in Keeney's value-focused thinking \citep[see][]{keeney1994creativity}, prioritizing values over available alternatives and fostering innovative solutions. They represent values hierarchically but may not capture the problem's full complexity. The lack of a standardized methodology can lead to information loss and attribute asymmetry \citep[see][]{jacobi2007quantifying,poyhonen1998notes} necessitating a critical application in decision-making.

Our approach integrates cognitive maps and value trees, key components of PSMs like Strategic options development and analysis \citep[SODA,][]{ackermann2001soda} and  Value-focused thinking \citep[VFT,][]{Keeney92}. Cognitive maps visualize stakeholders' perceptions and structure complex problems, while value trees explore potential solutions hierarchically. Combining these tools enhances problem structuring and solution generation in conflict transformation and management.


However, improving the process allowing to construct a value tree relevant for problem situation has not handle the problem how to design innovative solutions. For this purpose we need formal design theory.

\subsection{Design Theory: C-K Theory and P-KCP Methodology}\label{cktheory}

Design theory positions design at the core of professional practices, distinguishing it from natural sciences through its focus on creating preferred situations from existing ones \citep{simon1969sciences}. This foundational perspective views design problems as inherently ``wicked'', due to their complexity and the interconnectivity of solutions and issues \citep{simon1988science}. The interdisciplinary nature of design research, highlighted by \cite{chakrabarti2016anthology}, showcases the evolution of design theory over the last fifty years. This evolution, as outlined by \cite{bayazit2004investigating}), demonstrates a shift from systematic and rational processes to embracing disruptive innovations and creative reasoning, recognizing the need for interdisciplinary approaches and user involvement in solving complex real-world problems.

Modern design theory, particularly through the lens of C-K theory introduced by \cite{hatchuel2003new}, represents a significant advancement by providing a methodology for innovative design processes. This theory encourages the generation of novel alternatives by leveraging existing knowledge to explore the unknown, addressing the need for breakthrough innovations and the development of new expertise.

In the context of a specific knowledge domain (K-space), a "concept" refers to a proposition or group of propositions that typically describe attributes qualifying entities \citep[see][]{hatchuel2003new,elmquist2009sustainable}. The ``concept space'' (C-space) encompasses all concepts relative to K and is designed to be expandable to adapt to the dynamic nature of design and innovation. Co-evaluation of K-space and C-space emphasizes the creative design process. Any improvement in one space will lead to improvements in the other. C-K theory establishes that a C-space must have a tree structure, as it operates through partitions and inclusions, enabling the exploration and generation of new concepts or design objects.

The KCP (Knowledge, Concepts, Proposals) methodology has been proposed for collaborative design \citep{hatchuel2009ck,agogue2014introduction}. \cite{ferretti2019studying} emphasize the overlap between decision and design theory in decision-making, exploring their integration with policy studies and operational research to enhance policy design methods. \cite{pluchinotta2019design,pluchinotta2020integrating} further this understanding by introducing the P-KCP  methodology for public policy domains. \cite{pluchinotta2019design} proposes the P-KCP methodology which emphasizes participatory processes and stakeholder involvement. This approach is designed to enhance public decision-making by integrating design theory, specifically the C-K theory, to foster the generation of innovative policy alternatives.

\subsection{Water management}\label{watermgt}

Garrett Hardin's "The Tragedy of the Commons"  \citep[see][]{hardin1974tragedy} illustrates how individual self-interest in the use of shared resources leads to overexploitation and depletion, proposing that only external interventions or a shift in moral values can prevent such outcomes. Ostrom \citep[see][]{ostrom1990governing,ostrom1999coping} provides evidence that communities can effectively manage common-pool resources through self-organized governance structures, challenging Hardin's presumption of inevitable tragedy, but nothing is suggested on how to prevent or handle the inevitable conflicts arising in such settings. In the context of groundwater management in Tunisia, this paper adopts the participatory knowledge co-production process P-KCP to address these inherent conflicts, emphasizing the importance of engaging stakeholders \citep[see][]{molle2020comanagement,daniell2012co} in creating sustainable groundwater management strategies. This approach aims to bridge individual and collective interests, ensuring the equitable and sustainable utilization of groundwater resources.

In Tunisia, the critical role of groundwater for agriculture, coupled with its arid climate and the challenges of overexploitation, has been studied in literature  \citep[see][]{tringali2017insights,mekki2013management,mekki2017perceptions,saidi2013groundwater,dhaoui2022integration,soula2023evaluation}.

Groundwater in Tunisia is a critical resource, providing over  40 \% of the country's irrigation water and it has faced significant depletion due to overexploitation over the last three decades \citep[see][]{frija2014critical}. Despite the existence of an official groundwater management strategy, its enforcement is challenged by weak implementation capacities at the regional and local levels. This situation underscores the urgent need for improved governance and enforcement mechanisms to ensure the sustainable use and management of groundwater resources in Tunisia. In this context, the participatory approach highlighted in  study on Aousja Ghar El Melh \citep[see][]{hassenforder2024participatory} becomes particularly relevant, as in this paper will delve into how a participatory approach can address these challenges, fostering collaboration and stakeholder engagement as pivotal components for the aim of sustainable groundwater management in Tunisia. \citep[see][]{bouzidi2023sustainable} discusses groundwater overexploitation, emphasizing the limitations of regulatory approaches and the quest for new water sources. It proposes drawing lessons from three successful management experiences: borehole management, irrigated area management, and volumetric management.

Our study proposes an innovative and participatory approach by merging PSMs with the P-KCP methodology, aiming to confront the multifaceted challenges of groundwater management in Tunisia.


\subsection{Our proposal}\label{proposal}

This research explores the integration of PSMs with the P-KCP methodology and Concept-Knowledge (CK) theory, using cognitive maps and value trees as tools to generate new alternatives for collective groundwater management in Tunisia. This approach represents an advancement in the field by blending descriptive methods with theory-driven frameworks, aiming to improve and innovate policy design in water management.

In our research, we adapt an innovative approach similar to \cite{pluchinotta2020integrating} and \cite{pluchinotta2019design}
study, which integrates PSMs with C-K theory. While \cite{pluchinotta2020integrating}
utilized Fuzzy Cognitive Maps for stakeholder analysis, our study in Tunisia employs aggregated cognitive maps as a descriptive tool and further develops value trees as prescriptive tools derived from these maps.

More specifically we used the procedure introduced in \cite{tosunlu2023conflict} consisting in the following steps: \\
 - elicit cognitive maps for the relevant stakeholders involved in the problem situation through directed interviews; \\
 - transform the cognitive maps in ``value cognitive maps'' where all concepts present in the cognitive map become ``values''; \\
 - transform the value cognitive map in ``ends-means graph'' resolving the problem of cycles; \\
 - transform the ends-means graph in value trees (satisfying the unique conditions of a tree structure). \\
This process has been designed for conflict transformation purposes with the aim of establishing common ground for discussion when manifest conflicts (possibly violent) keep the stakeholders distant.

However, in the present case study we introduced a light modification of the procedure and added one further step to it. In absence of a manifest (possibly violent) conflict and given the cooperative attitude among the stakeholders (although not conflict free) we decided to unify the different cognitive maps to a single one resuming how the problem situation was perceived by the different stakeholders and using this as a basis for the further development of the procedure. The further step introduced (and tested) for this case study, has gone beyond the construction of the value tree (representing the decision making drivers of the group) and derives a ``concept tree'' where more attributes and their combinations can be tested in order to figure new potential actions or to identify otherwise unforeseeable consequences which could become potential hard constraints for policy design. In other terms, using design theory terminology, we used the value tree as a ``knowledge space'' (what we know about the actual decision making drivers of the stakeholders) from which create a ``concept space'' (what could be decision making drivers for the stakeholders in case a policy design is tested).

Typically, creating a concept tree demands considerable time and relies heavily on the facilitator's personal insight. This aligns with literature \citep{hatchuel2009design,hatchuel2004ck}, which underscore the challenges of navigating and expanding the C-space. Moreover, in the literature \citep[see][]{kazakcci2009formalization} the elaboration of the concept space in design involves a complex process of partitioning concepts and validating them through K-validation, which assesses the feasibility of design propositions based on existing knowledge. The process tested in this experiment (see Section \ref{processresults}) consisted in incorporating concepts little discussed during the workshops (unchecked consensus) together with ``disruptions'' \citep[see][]{considine2018thinking}, including questioning the explicit absence of subsets of values.

The whole experiment has been conducted through interviews and 3 workshops implying local farmers, the administration, the researchers and other relevant stakeholders. The problem situation of the case study is presented in Section \ref{case study}, while the workshops and their results are presented in Section \ref{processresults}.

\section{The Area: Case Study}\label{case study}

In order to present our framework, we discuss a real-world case study, concerning water management in Tunisia. The experiment consisted in conducting three participatory policy design workshops besides the whole preparatory activities. For a more detailed description of the problem situation the reader can see \cite{chrii2022exploration}.

Geographically, the Limaoua area is strategically positioned within the Gabès governorate in Southern Tunisia, bordered by the city of Gabès, Zeuss stream, the sea, and the Gabès Matmata airport. It is part of the Jeffara aquifer system, incorporating the Gabès North, Gabès South, and El Hamma Henchou aquifers \citep[see][]{vernoux2019scenarios}. The area significantly depends on the Gabès South aquifer's lower senonian carbonate aquifer, found at depths of 60 to 250 meters. To address water overexploitation, the administration introduced a "safeguard zone" in 2017 to regulate drilling depth to enhance sustainable water management \citep[see][]{chrii2022exploration}.

\begin{center}
 \begin{figure}
  \scalebox{0.6}{\includegraphics{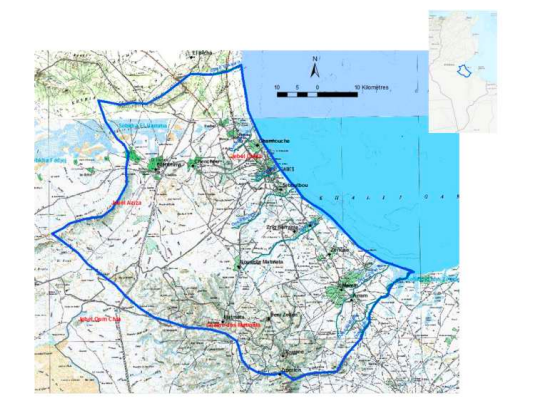}}
    \caption{Gabes area-CRDA, 2021}\label{Gabes}
 \end{figure}
\end{center}

The problem arises from the over-exploitation of the Gabès South aquifer, with annual withdrawals estimated at 47 million cubic meters against a recharge of 36 million cubic meters per year (CRDA, 2016). This overuse, coupled with low annual rainfall of 180mm and increasing water points post-2011, led to a comprehensive 2021 inventory identifying 1597 water points, highlighting a mix of public, private, simple, and illicit wells. The escalation in water points, especially following the 2011 Revolution and the safeguard zone's establishment, aligns with periods of drought, exacerbating water scarcity concerns \citep[see][]{soula2021water}.

Agricultural activities, particularly arboriculture, dominate water use in Limaoua, with around 630 farmers operating sizable plots and benefiting from fertile soil, energy access, and infrastructure. However, the influx of new residents and resultant well drilling have strained water resources, leading to aquifer depletion concerns. The 2017 safeguard zone's establishment, amid drought, heightened tensions due to increased illicit drilling and administrative enforcement challenges. This situation risks further complications from potential saltwater intrusion and stricter regulations if the area becomes a ``prohibited area''.

In the 2000s, efforts were made to establish an Agricultural Development Group (GDA) for collective groundwater management in Limaoua, aimed at regulating water use through shared, high-capacity wells. GDAs facilitate administration control over water volumes used, contrasting with the challenge of monitoring numerous private wells as described by \cite{chrii2022exploration}. A successful example of GDA management is Bsissi Oued El Akarit, indicating potential benefits for Limaoua, despite challenges in enforcement and the unique context of private irrigation management.


Inspired by Bsissi's GDA success,  the Regional Commission for Agricultural Development (CRDA) of Gabès sought to implement a similar model in Limaoua during the 2000s but faced challenges due to a lack of leadership, engagement and resources. As water demand has increased, reconsideration of this initiative is underway. However, the dynamics between CRDA and farmers have significantly changed in the current post-revolution context, making the direct application of previous strategies more complex \citep[see][]{frija2016gestion,molle2020comanagement}.

For this reason, the CRDA sought the support of Center for International Cooperation in Agricultural Research for Development (CIRAD) in Tunisia, which specializes in supporting participatory natural resource management. This led to the decision that it could be suitable to experiment with new policy design tools. As a result, CIRAD sought the help of the University of Paris Dauphine, and a research team was formed in collaboration with the Australian National University (ANU), University College London, and the French National Centre for Scientific Research (CNRS).

\section{Method}\label{method}

In our methodological approach, we employed the P-KCP (K for knowledge, C for concepts and P for proposals) methodology, a participatory tool for the innovative design of policy alternatives as introduced by \cite{pluchinotta2019design}. The P-KCP methodology, effective in generating innovative policy solutions and enhancing stakeholder collaboration, was particularly suitable for our study's focus on groundwater management in Tunisia. To structure our activities, we categorized them into three main stages consistent with the P-KCP methodology. These stages were designed to facilitate a generative and participatory process, allowing us to explore a range of policy alternatives. This approach not only aligns with the P-KCP's emphasis on creating novel policy alternatives but also addresses the specific challenges encountered in the public policy domain of Tunisia.

The P-KCP methodology consists of four phases as described in \cite{pluchinotta2019design}; for the purpose of the research, we will focus on the first three phases. The roadmap of the experiment is summarised in Table \ref{tab:roadmap}.

\begin{table}
\footnotesize
    \centering
    \begin{tabular}{lll} \hline
        \textbf{Policy-Definition} &
        \textbf{Policy-Knowledge} &
        \textbf{Policy-Concepts Generation} \\ \hline
        \parbox[t]{4cm}{\begin{itemize}
            \item Conduct preliminary interviews with stakeholders to understand diverse perspectives.
            \item Build individual cognitive maps for each stakeholder from the interview transcripts.
            \item Organize Workshop-1 to share and gather knowledge about water resources in Limaoua.
            \end{itemize}} &
        \parbox[t]{4cm}{\begin{itemize}
            \item Finalize the aggregated cognitive map, integrating insights from stakeholder interactions from individual cognitive maps.
            \item Facilitate Workshop-2, focusing on identifying the shared concern through a participatory approach.
            \end{itemize}} &
        \parbox[t]{4cm}{\begin{itemize}
            \item Develop a value tree from the cognitive map, using it as a benchmark for the concept tree.
            \item Facilitate Workshop-3 aimed at identifying and prioritizing solutions for collective groundwater governance by creating a "concept tree" that addresses the shared concern.
            \item Enrich the concept tree with findings from Workshop-3 and inputs from various stakeholders.
            \item Hold an expert workshop to refine and finalize the concept tree, incorporating comprehensive policy alternatives.
            \end{itemize} } \\ \hline
    \end{tabular}
    \caption{The experiment Road Map}
    \label{tab:roadmap}
\end{table}
\subsection{Policy–Definition Phase (P–D Phase)}\label{pdphase}

In the P-D Phase of our study, there were three main objectives \citep[see][]{pluchinotta2019design}: (i) to gather and analyze existing knowledge on water management in Limaoua, forming a foundational understanding; (ii) to understand stakeholders characteristics, focusing on their objectives and values; (iii), to gain an initial understanding of the problem from different stakeholder perspectives.





\subsection{Policy–Knowledge Phase (P–K Phase)}\label{pkpahse}

The P-K Phase of our study, as defined by \cite{pluchinotta2019design}, aimed to establish a collective foundation for policy development. This phase was expected to produce several key outcomes: (i) a comprehensive summary of state-of-the-art knowledge on the case study and policy issue, (ii) an enhanced and detailed stakeholder analysis, (iii) a common problem formulation incorporating individual viewpoints, and (iv) the identification of the dominant design in traditional policy alternatives through a preliminary C-tree model.





\subsection{Policy–Concepts Generation Phase (P–C Phase)}\label{pcphase}

In the P-C Phase, we embarked on a collaborative journey with stakeholders to craft innovative policy alternatives within the Concept-Knowledge (C-K) theory. This process began by establishing a shared problem understanding, setting a solid foundation for our explorative endeavor. We then introduced the Concept tree (C-tree), a tool that served as a guide for our collective exploration of policy options. Participants, divided into diverse groups, engaged in rich discussions, proposing expansions to the C-tree. This collaborative effort aimed at reaching a consensus on innovative alternatives, culminating in a reflective discussion that solidified our findings and insights. Through this participatory and generative workshop, we utilized the C-space to visualize and map out a range of policy alternatives, fostering an environment of creativity and collective problem-solving.

\section{Process and Results}\label{processresults}

This section describes the methodology applied and the associated results in the case study.

\subsection{Policy-Definition Phase (P-D Phase)}\label{pd2}




The preliminary interviews and cognitive map, in conjunction with Workshop-1, constitute the critical components of the P-D Phase in our process. The initial phase of this research entailed a series of preparatory activities to lay a groundwork.
We prioritized establishing direct dialogue with local key stakeholders, integral to decision-making in groundwater management. Stakeholder selection aimed to include a variety of viewpoints and interests to provide a well-rounded perspective. Additionally, an interview framework was devised to drive substantive conversations during the cognitive mapping stage.

For Workshop-1 farmers were divided in two groups based on the size of their land holdings, recognizing that land size often signifies differing interests and concerns within the agricultural community \citep[see][]{frija2016farmers}. This approach allowed for more focused discussions, ensuring that the unique challenges and perspectives of both small and large farmers were adequately addressed. 
Participants were exposed to strategies and outcomes from Bsissi's approach to groundwater governance \citep[see][]{minoia2008social}, offering a rich source of knowledge for consideration and application in the broader context of groundwater management in Tunisia.


With the completion of preliminary interviews, the development of cognitive maps for each stakeholder (or category of), and the conduct of Workshop-1, we have successfully concluded the P-D Phase of our process. 

\subsection{Policy–Knowledge Phase (P–K Phase)}\label{pk2}

Transitioning into the (P-K) Phase, our study embarked on a strategic path to further engage with and understand the complexities of groundwater management in Tunisia.
This phase was marked by two activities: establishing an aggregated cognitive map and identifying the shared concern among the stakeholders Workshop-2.  

\subsubsection{Aggregated cognitive map}\label{aggregcm}

The aggregated cognitive map aims to achieve a unified problem formulation that integrates individual perspectives and summarizes current knowledge on the case study.

The cognitive mapping process in our study was carried out in stages. Initially, we conducted individual mapping exercises with each stakeholder group, capturing their distinct perspectives and concerns. These individual maps offered insights into their knowledge and perceptions about the groundwater management issue. Following this, we created aggregated cognitive maps, merging these individual perspectives. This amalgamation facilitated the identification of both commonalities and variances among the groups.
Importantly, during the aggregation process, we ensured no loss of information from the individual maps as we recorded each concept and utilized them in the subsequent development of a concept tree, further enhancing our policy development process.


When aggregating individual maps, it is necessary to first mention the fundamental node. The fundamental node can be referred to as the main goal/purpose in a map \citep[see][]{tosunlu2023conflict}. The existence of a common fundamental node makes aggregation possible.
Since each stakeholder's fundamental node is related to agriculture, common nodes that influence this node were selected and merged into a single map. The reason for merging the stakeholders' individual maps is that there is no declared conflict yet, and we want to focus on similarities rather than differences.

\begin{sidewaysfigure}
\centering
  \includegraphics[width=15cm]{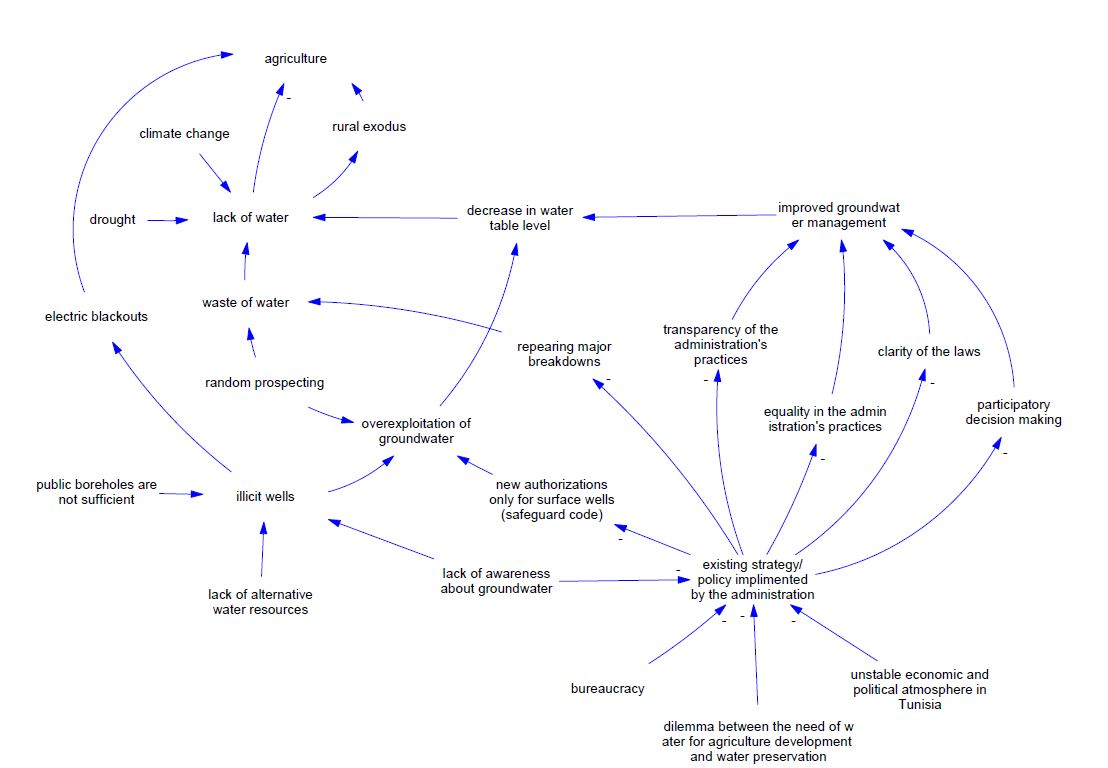}
   \caption{The aggregated cognitive map}\label{cmtunis}
\end{sidewaysfigure}
Stakeholders' reactions to the aggregated cognitive map were largely favorable, particularly regarding the general map. Small farmers, who sometimes held distinct views from other stakeholders, also found common ground, possibly due to the focus on similarities rather than disparities. For more details about the stakeholder categorization and details of the aggregated cognitive map, see the Supplementary Material.

The cognitive map was instrumental in achieving the P-K phase's objectives by structuring stakeholder insights and fostering a common understanding of water management challenges.
This structured representation not only supported consensus-building among diverse stakeholder groups but also highlighted the interconnectedness of issues, demonstrating its value in enhancing collaborative problem solving efforts within the framework.

\subsubsection{Definition of the shared concern}\label{shared}

In Workshop-2 18 stakeholders from diverse backgrounds convened to collaboratively explore groundwater management challenges divided in two groups. This included representatives from CRDA, CTV, GDA, and both large farmers and small farmers, emphasizing the importance of finding common ground. By focusing on similarities rather than differences, facilitators steered the discussions towards unity and collaborative solutions, enhancing the group's collective approach to addressing the complexities of water resource management in Limaoua. 

To identify the shared concern, each group constructed a sentence summarizing its objective. The first group chose 'finding solutions to organize farmers within a guaranteeing authority, ensuring sustainable agriculture, income improvement, water control and management, and access to subsidies'. The second group emphasized the need for good relationships between stakeholders, reasonable water management, and the use of irrigation techniques. Their objective was to find solutions for good water management in a participatory framework. Both sentences were presented to all participants, and the first sentence was chosen as it was considered to encompass all their goals. Importantly, stakeholders reached a consensus on the fundamental concept (node).

Utilizing  the aggregated cognitive map within the P-KCP methodology significantly enriched the outcomes of Workshop-2, as highlighted by the stakeholders. The use of the cognitive maps made the workshop more focused and productive, facilitated a more structured exchange of perspectives, and allowed to identify the shared concern. 

\subsection{Policy-Concepts Generation Phase}\label{pc2}

Transitioning into the P-C Phase, our methodology encompassed constructing a Value Tree, organising Workshop-3, and constructing two Concept trees with an expert workshop aimed at advancing towards actionable groundwater management strategies.

The steps of this process can be summarised as follows.\\
1. Generating a Value Tree out of the aggregated cognitive map. \\
2. Conduct Workshop-3 in order to identify a first set of alternatives meeting the consensus of the stakeholders. \\
3. Construct a concept space based upon the value tree and the discussions in Workshop-3. \\
4. Refine the concept space through a restricted experts' workshop.

\subsubsection{Value Tree} \label{sssectionvt}

\begin{sidewaysfigure}
\begin{center}
    \includegraphics[width=17cm]{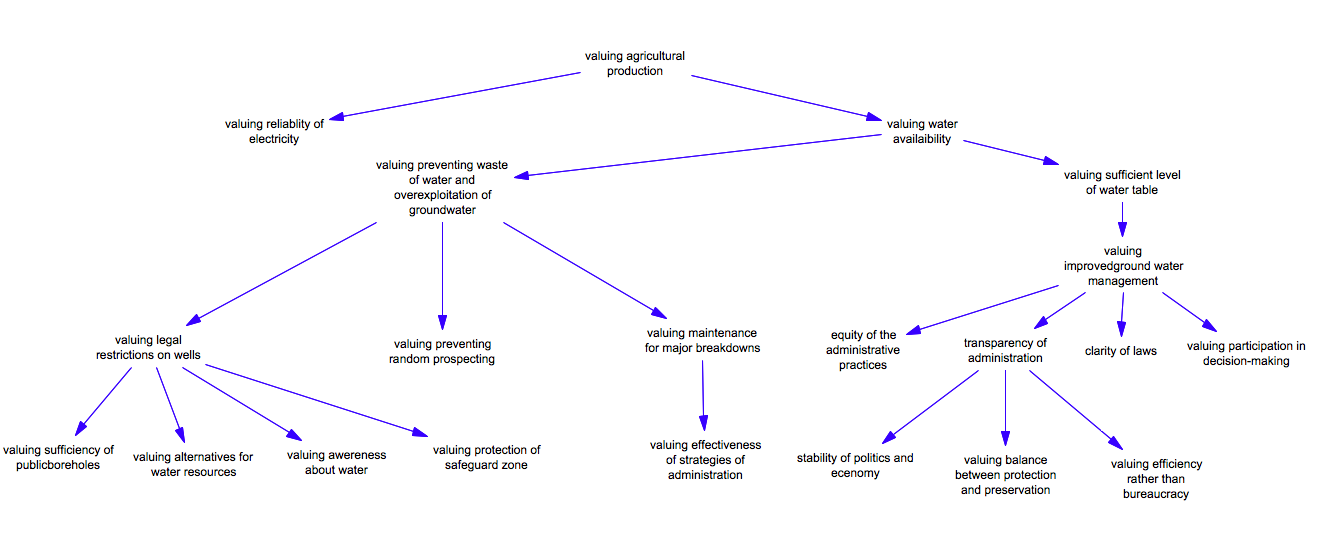}
    \caption{The value tree out of aggregated cognitive map}\label{vttunis}
    \end{center}
\end{sidewaysfigure}

The construction of the value tree followed the steps already presented in \cite{tosunlu2023conflict}. The result is visible in Figure \ref{vttunis}. This value tree has not been explicitly presented to the stakeholders, but has been used for two purposes. \\
1. Validate the proposal emerging from Workshop-3, ensuring that these were compatible with the values representing the consensus established among the stakeholders. \\
2. Use it as a basis for constructing, after the Workshop, the concept space to be used for further innovative design of alternatives.

The interested reader can check the whole process of constructing and validating the value tree in the Supplementary Material. For the purpose of this paper, anticipating the discussion in Section \ref{sssectionwrks3}, we emphasize that
the value tree effectively captures the innovative solutions identified by stakeholders, with each solution represented or derivable within its structure. Non-corruption is addressed through "equity in administrative process" and "transparency of administration." Regulating illicit drilling is captured by "valuing preventing random prospecting" and "valuing legal restrictions on wells." Debt support for farmers is linked to "valuing efficiency over bureaucracy." Training is highlighted as "valuing awareness about water." Understanding GDA financing resources is covered under "clarity of laws." Collective authorization for small farmers is derivable from "valuing alternatives for water resources." Revising laws for GDA operations is tied to "valuing participation in decision-making" and "transparency of administration."




\subsubsection{Workshop-3}\label{sssectionwrks3}

Workshop-3 convened 33 participants, including two facilitators, three authors of this paper, and stakeholders from various sectors, including 18 farmers. Overall, Workshop-3 successfully generated alternatives and outlined a roadmap for creating and implementing the GDA, aimed at collectively managing groundwater and addressing shared concerns in the region. The aggregated cognitive map played a crucial role in the Policy-Concepts Generation Phase by offering a structured overview of the situation, highlighting stakeholder perspectives, and facilitating a comprehensive review.

Several non exclusive actions have been identified as necessary steps for a sustainable water management policy. These were: \\
 1. Fight against corruption in the administration. \\
 2. Regulate illicit drilling. \\
 3. Renegotiate farmers' debts. \\
 4. Training for farming new sustainable crops. \\
 5. Creating a financially viable GDA. \\
 6. Establish a collective authorisation scheme for small farmers. \\
 7. Revise the existing laws regulating the GDAs.

Among these the first 5 actions were inspired by existing water management plans, essentially around the establishment of a GDA. The last two actions were specifically introduced during the workshop as a result, on the one side, of a solid group of small farmers and, on the other side, of the accumulated experience of other GDAs management experiences.

As already mentioned the elaboration of these actions has been checked against the value tree. Actually the seven actions can be seen in the value tree as follows: \\
1-Non-corruption: The value tree encompasses this through "equity in administrative process" and "transparency of administration," offering a nuanced approach to addressing corruption. \\
2-Regulating illicit drilling: This is addressed by "valuing preventing random prospecting" and "valuing legal restrictions on wells," capturing the essence of regulation within the value tree.\\
3-Debt support for farmers: Represented as a means towards "valuing efficiency over bureaucracy," aligning financial support with broader efficiency values.\\
4-Training: Included in the value tree as "valuing awareness about water," highlighting educational aspects as crucial for sustainable management.\\
5-Understanding GDA financing resources: Covered under "clarity of laws" in the value tree, emphasizing the importance of clear, accessible financial frameworks.\\
6-Collective authorization for Small farmers: This solution concept, while new, can be derived from "valuing alternatives for water resources," suggesting a broader inclusivity in resource access.\\
7-Revising laws for GDA operations: Tied to "valuing participation in decision-making" and "transparency of administration," indicating a push for more inclusive and transparent governance structures.\\

\subsubsection{Initial Concept tree}\label{sssectioct1}

\begin{sidewaysfigure}
\begin{center}
    \includegraphics[width=15cm]{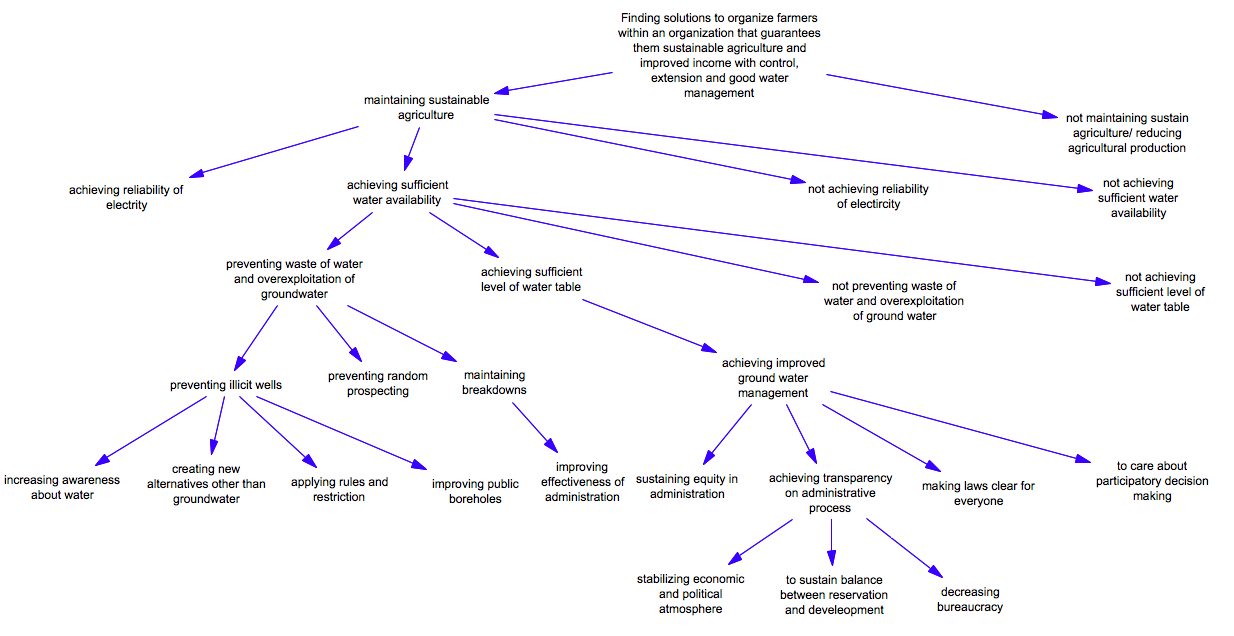}
     \caption{Initial concept tree- out of the value tree}\label{cttunis}
    \end{center}
\end{sidewaysfigure}

Following the proposal described in Section \ref{proposal} and having validated the value tree in Workshop-3 as a knowledge space (from which the the first actionable ideas have been decided) we constructed a first concept space incorporating all relevant, yet previously not utilized, concepts from the preliminary interviews and cognitive maps, thus minimising risks of information loss \citep[see][]{jacobi2007quantifying,poyhonen2001behavioral}. This enrichment process significantly enhanced the depth and comprehensiveness of the concept tree, ensuring a broader spectrum of solutions was considered and the final concept tree was constructed. The result in shown in Figure \ref{cttunis}.

This was done essentially for research purposes: testing the capability of the concept space to be the driving structure for the alternatives generative process.

The initial concept tree, rooted in the hierarchical structure, begins with the overarching shared concern of devising strategies to organize farmers within a framework that promises sustainable agriculture and enhanced income through effective control and optimal water management. This aligns with the fundamental value of "valuing agricultural production" identified in the value tree. The tree first starts with  "sustainable agriculture" with its negation, trying to provoke solutions that could boost farmer income without relying solely on agriculture.



The next step will integrate findings from the interview guides not previously included since the aggregated cognitive map only focused upon commonalities.


\subsection{Final concept tree with experts expansion}

In order to address potential attribute loss due to  oversimplification and the tree's restrictive nature \citep[see][]{von1987value} we organised a restricted workshop among experts and researchers aiming at refining the initial concept space.

\begin{sidewaysfigure}
\begin{center}
    \includegraphics[width=20cm]{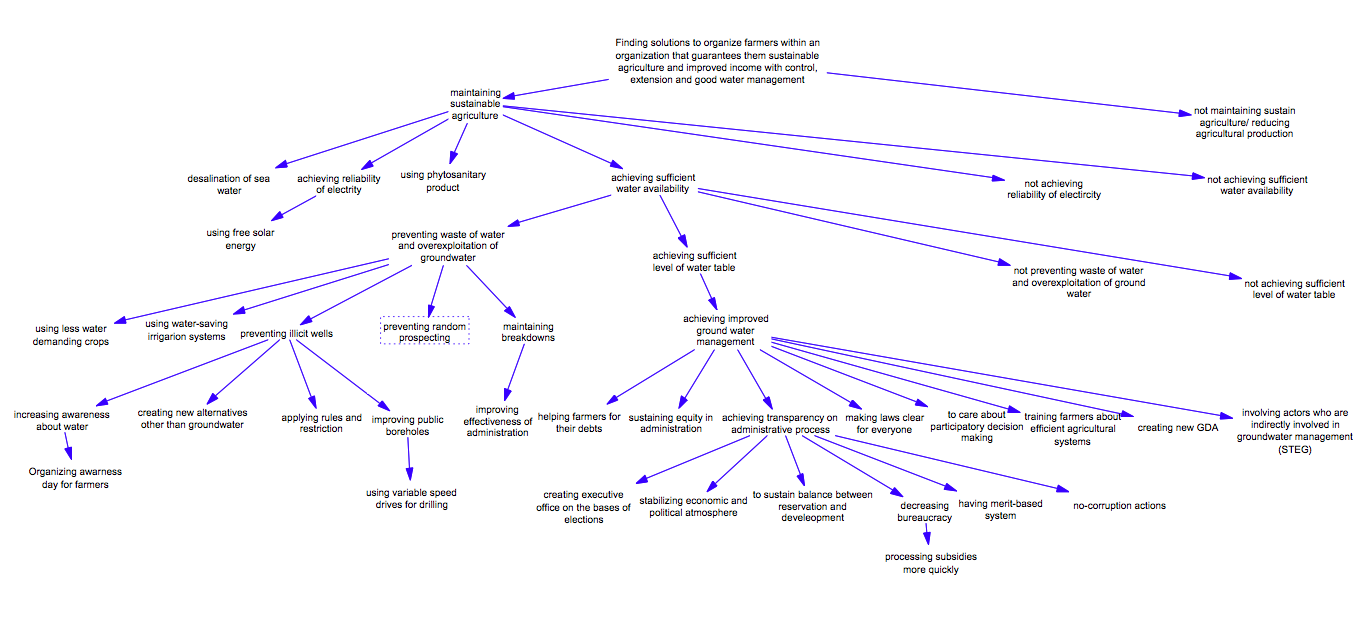}
     \caption{Final concept tree}\label{fcttunis}
    \end{center}
\end{sidewaysfigure}

This step addresses the fact that the value tree, designed to highlight shared concerns and values, might inadvertently omit certain stakeholder-specific concepts. By integrating these neglected concepts, we enhance the comprehensiveness of the concept tree, ensuring it represents a fuller spectrum of stakeholder perspectives and insights. This inclusive approach rectifies the potential loss of valuable insights through aggregation and subsequent filtering processes, thereby enriching the final policy design framework with broader considerations and alternatives. The result is visible in Figure \ref{fcttunis}.

The concept tree was expanded to include new concepts identified during Workshop-3, reflecting a broader spectrum of stakeholder insights and priorities. It now incorporates discussions on phytosanitary products, seawater desalination, the introduction of novel crops, and a shift towards more water-efficient crop practices. A notable divergence was observed between Small farmers and Large farmers regarding the issuance of new well authorizations, highlighting the complexity of groundwater management. This diversity in viewpoints underscores the importance of considering the varied interests and concerns of different stakeholders in developing sustainable management solutions.

\begin{sidewaysfigure}
    \includegraphics[width=20cm]{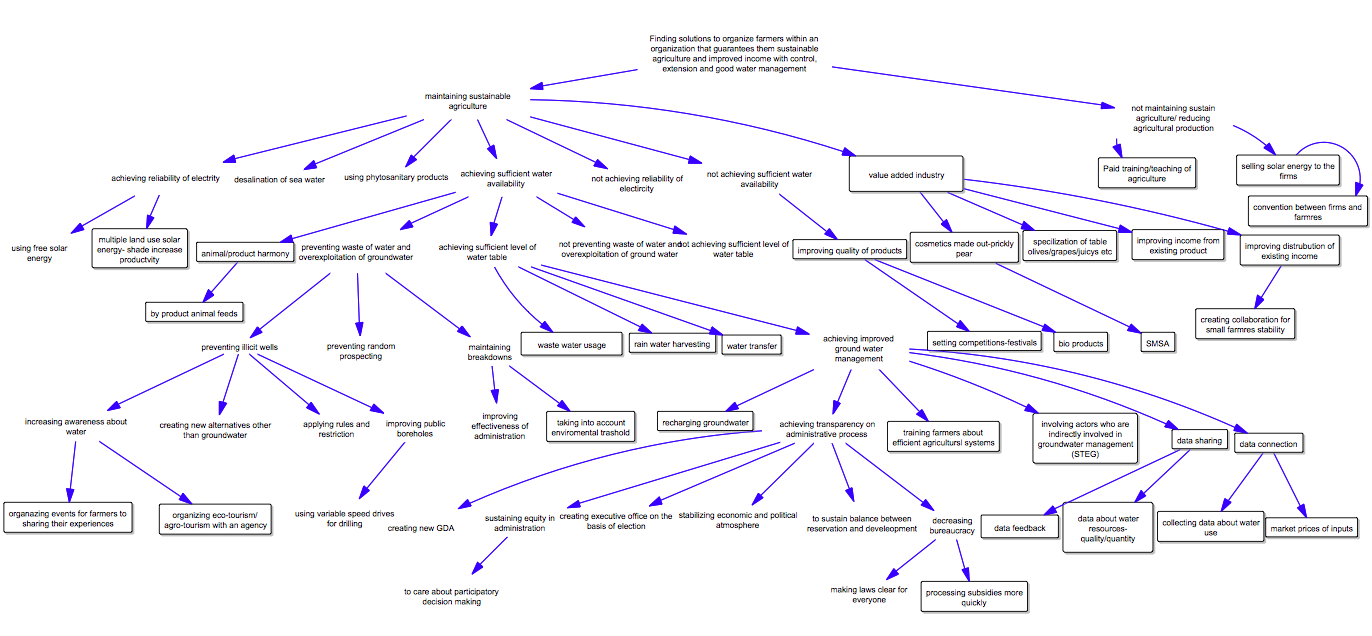}
     \caption{Final concept tree with experts expansion}\label{fctetunis}
\end{sidewaysfigure}

\section{Discussion}

In the following we briefly discuss the experiment and the lessons learned from this action research project.

\subsection{Methodology}

In our research, we employed the P-KCP methodology in a real-world case study set in the Limaoua region, adopting an approach similar to that of \cite{pluchinotta2019design} but diverging by integrating cognitive mapping and value trees. 
We advocate for the integration of a value tree (out of the cognitive map) as the foundational element when embarking on the creation of a concept tree. 

Our methodology sought to stimulate stakeholders' problem-solving capacities using cognitive maps produced through an interview guide. Subsequently, we aimed to gain a deeper insight into the commonalities and disparities between stakeholders by amalgamating these individual cognitive maps into one comprehensive representation. We derived a value tree from the aggregated map, 
thus gaining two advantages: \\
 - the assurance that ideas and proposals arising from the participants would meet the consensual values of the group; \\
 - a knowledge space from which start the generative process for creating innovative solutions.

We contend that incorporating a value tree not only simplifies the tasks for stakeholders and facilitators but also sparks creativity in the development of a concept tree. As such, we propose that the existing C-K literature recognizes the value tree as a fundamental and structured component of the concept tree.

\subsection{Participation}

An important lesson learned from this experiment has been recognising the importance of a ``model driven participation''. With this term we intend the use of a formal methodology on how participation should be conducted beyond brainstorming or just general discussion. If participation is supposed to produce any usable results then it has to be driven by some model. In this experiment we conducted 3 workshops. Workshop-1 primarily involved the sharing of individual concerns and varied knowledge. In  Workshop-2, participants gained a deeper understanding of the issue through the presentation of the aggregated cognitive map (CMs). Workshop-2 facilitated a deeper understanding of the interconnected nature of the problems, encouraging more detailed discussions and a focus on finding collective solutions. Finally Workshop-3 has been essentially driven by the value tree derived from the aggregated cognitive map. Overall we had a much more effective and efficient participative process, despite the many problems participation implies.\footnote{The reader should note that the whole experiment and the workshops have been conducted during the COVID pandemic imposing the well known restrictions.}

\subsection{Practical results}

From a practical point of view (the one of the Tunisian administration and of the farmers) the workshops (and the whole process) allowed to construct a legitimated set of proposals as far as the groundwater management problem of the Limaoua area is concerned. The proposal to which Workshop-3 arrived were reasonable, convincing for the whole set of participating stakeholders and actionable from the point of view of the administration. This was not at all obvious at the beginning of the process given the differences, both substantial and perceived, among the different stakeholders. From the client's point of view the consensus reached about the establishment of a GDA was a success because gained the legitimacy from the farmers (who should implement it). At the time of the end of the workshop there was a clear operational direction to follow.\footnote{The unstable political situation in Tunisia did not allow to have further knowledge about the practical implementation of the GDA in Limaoua.}

\subsection{Research results}

The experiment confirmed the validity and interest of establishing cognitive maps as the first step in constructing a descriptive model of what the problem is for each stakeholder and check how distant are these mental (and material) representations, which also allows a rough estimation of potential conflict escalations. It also confirmed the utility of deriving a more prescriptive representation of the decision making driving structure: the value tree. This provides assurance that the proposal emerging from the participative process are legitimate with respect to the collective values of the group. The experiment confirmed the intuition consisting in using the value tree as a knowledge space from which start a design (creative) process formalised as the construction of a concepet space (coming in the form of a concept tree).

Comparing the three figures (Figures \ref{vttunis}, \ref{cttunis} and \ref{fcttunis}) we can observe how the concept tree allowed to incorporate all major issues valued during the design process. Our concept tree effectively encapsulates a multifaceted typology of water management strategies, showcasing innovative solutions for sustainable resource use in Tunisia \citep[see][]{hassenforder2023accompagner}. It integrates key elements such as Supply Management, highlighted by initiatives like rainwater harvesting and desalination of seawater, and Demand Management, where we address the prevention of water waste and overexploitation. Alternative Resources are represented through the use of free solar energy.

Further, the concept tree incorporates Aquifer Recharge techniques to ensure long-term water sustainability, while the Water Accounting principles are mirrored in our data sharing and data connection nodes, ensuring transparency and efficient resource allocation. The concepts of Drilling Operations Management are evident in our efforts to prevent illicit wells, fostering responsible water extraction. Through these examples, our concept tree provides a comprehensive blueprint that aligns with the recognized typologies, directing us towards a holistic approach to water management in Tunisia.
In adapting the comprehensive measures from the Groundwater Catalogue(https://www.groundwatercatalogue.org/measures) to our paper, we can integrate management solutions that correspond with the innovative approaches outlined in our concept tree. For instance, our concept tree's inclusion of 'rainwater harvesting' and 'desalination of sea water' aligns with the Catalogue's supply measures. Additionally, 'preventing waste of water' and 'improving public administration effectiveness' echo the demand and protection measures, emphasizing stakeholder involvement and policy dialogue. This synergy underscores our commitment to a multidimensional approach to water management in Tunisia.

We also had positive feedback as far as the different steps tested for the construction of the concept tree are concerned. \\
1. Contrasting the nodes of the value tree allowed to provoke the reaction of the stakeholders: understanding why ``not supporting agriculture'' is not a viable option makes clear which are the profound reasons for which it is necessary to pursue a viable groundwater management policy for agricultural development. \\
2. Incorporating experts' concerns and precise suggestions allowed to add potential solutions otherwise concealed by the consensual process of the workshops activities. \\
3. Incorporating single stakeholders experience and knowledge allowed to emerge critical aspects otherwise impossible to grasp. An interesting case concerns the consequences of using solar panels for producing electricity. The massive diffusion of this technology allowed producing cheap electricity encouraging more water pumping and more overexploitation of the aquifer. We now know that these energy sources need to be regulated as for the rest if a sustainable water management policy has to be established.

\subsection{Limitations}

Although our experiment has been a success both from the action point of view (suggest legitimated and actionable solutions for the sustainable groundwater management problem at the Limaoua area) and the research point of view (establish a whole process from the construction of cognitive maps of the stakeholders to the construction of a concept space supporting the design of innovative solutions), we have to recognise a number of shortcomings and limitations to be further analysed in a research perspective.

\begin{itemize}
    \item From a practical point of view, Workshop-3 has been less ``innovative'' than expected. There are several reasons for that (see the following bullet), but the fact that a substantial part of the stakeholders were committed to the establishment of the GDA limited the discussions to how this should be successfully implemented rather than exploring other solutions. On the other side, the creation of the GDA is innovative with respect to the existing practices and definitely committed to sustainability.
    \item Besides the difficulties related to the conjuncture (pandemic, political instability, etc.) the fact that the experiment has been conducted with participants having extremely different cultural, scientific and linguistic backgrounds has been an important limitation to the innovation capability of the whole experiment. The workshop facilitators were not at all familiar to design theory and decision support, but where native Arab speakers, while the researchers, experts in their domain, did not speak at all Arab, the whole experiment being conducted in this language. These are aspects which need to be considered better and in time when such experiments are designed.
    \item The construction of the concept space requires a dedicated workshop which we did not have the time and the resources to organise. The experts' workshop did help, but is an extra and cannot substitute the workshop we could not do.
    \item The generative process of solutions deriving from the concept space needs to be further formalised and developped. The same applies for the construction of the concept space itself.
\end{itemize}

\section{Conclusion}

In this study, we addressed the urgent need for innovative groundwater management strategies in Tunisia by integrating the P-KCP methodology with cognitive mapping and value trees. This approach not only facilitated a comprehensive understanding of stakeholder perspectives but also fostered the development of actionable and consensus-driven policy solutions. Our findings demonstrate the potential of this integrative method to enhance participatory groundwater management, making a significant contribution to both theoretical frameworks and practical applications. While our study provides a promising direction, further research is needed to explore its adaptability in varied contexts and to refine the tools used.

\bibliographystyle{apalike}
\bibliography{berkay2}

\begin{thebibliography}{}

\bibitem[Ackermann, 2012]{ackermann2012problem}
Ackermann, F. (2012).
\newblock Problem structuring methods ‘in the dock’: Arguing the case for
  {Soft OR}.
\newblock {\em European Journal of Operational Research}, 219(3):652--658.

\bibitem[Ackermann et~al., 2001]{ackermann2001soda}
Ackermann, F., Eden, C., Rosenhead, J., and Mingers, J. (2001).
\newblock {SODA}-journey making and mapping in practice.
\newblock In Rosenhead, J. and Mingers, J., editors, {\em Rational analysis in
  a problematic world revisited}, pages 43--61. J. Wiley, New York.

\bibitem[Ackoff, 1979a]{Ackoff79a}
Ackoff, R. (1979a).
\newblock The future of operational research is past.
\newblock {\em Journal of Operational Research Society}, 30:93--104.

\bibitem[Ackoff, 1979b]{Ackoff79b}
Ackoff, R. (1979b).
\newblock Resurrecting the future of operational research.
\newblock {\em Journal of the Operational Research Society}, 30:189--199.

\bibitem[Agogu{\'e} et~al., 2014]{agogue2014introduction}
Agogu{\'e}, M., Hooge, S., Arnoux, F., and Brown, I. (2014).
\newblock {\em An introduction to innovative design-elements and applications
  of {CK} theory}.
\newblock Presses de Mines, Paris.

\bibitem[Bayazit, 2004]{bayazit2004investigating}
Bayazit, N. (2004).
\newblock Investigating design: A review of forty years of design research.
\newblock {\em Design issues}, 20(1):16--29.

\bibitem[Bouzidi et~al., 2023]{bouzidi2023sustainable}
Bouzidi, Z., Faysse, N., Mekki, I., Ferchichi, I., Hassenforder, E., and
  Rinaudo, J.-D. (2023).
\newblock Sustainable management of groundwater resources in {Morocco} and
  {Tunisia}: how existing cases of functional groundwater management can help
  thinking about local solutions?
\newblock {\em Alternatives Rurales}, 9:on--line.

\bibitem[Chakrabarti and Blessing, 2016]{chakrabarti2016anthology}
Chakrabarti, A. and Blessing, L.~T. (2016).
\newblock {\em Anthology of Theories and Models of Design}.
\newblock Springer.

\bibitem[Chrii, 2022]{chrii2022exploration}
Chrii, S. (2022).
\newblock {\em Exploration d'options pour la gestion collective des eaux
  souterraines: Cas de Limaoua, Gab{\`e}s Sud}.
\newblock PhD thesis, INAT.

\bibitem[Colorni and Tsoukiàs, 2020]{ColorniTsoukias2020}
Colorni, A. and Tsoukiàs, A. (2020).
\newblock Designing alternatives for decision problems.
\newblock {\em Journal of Multi-Criteria Decision Analysis}, 27:150 -- 158.

\bibitem[Considine, 2012]{considine2018thinking}
Considine, M. (2012).
\newblock Thinking outside the box? {Applying} design theory to public policy.
\newblock {\em Politics and Policy}, 40:704--724.

\bibitem[Daniell, 2012]{daniell2012co}
Daniell, K.~A. (2012).
\newblock {\em Co-engineering and participatory water management:
  organisational challenges for water governance}.
\newblock Cambridge University Press.

\bibitem[Dhaoui et~al., 2022]{dhaoui2022integration}
Dhaoui, O., Antunes, I. M. H.~R., Agoubi, B., and Kharroubi, A. (2022).
\newblock Integration of water contamination indicators and vulnerability
  indices on groundwater management in menzel habib area, south-eastern
  tunisia.
\newblock {\em Environmental Research}, 205:112491.

\bibitem[Eden, 1988]{eden1988cognitive}
Eden, C. (1988).
\newblock Cognitive mapping.
\newblock {\em European Journal of Operational Research}, 36(1):1--13.

\bibitem[Eden, 1994]{eden1994cognitive}
Eden, C. (1994).
\newblock Cognitive mapping and problem structuring for system dynamics model
  building.
\newblock {\em System dynamics review}, 10(2-3):257--276.

\bibitem[Eden and Ackermann, 2004]{eden2004cognitive}
Eden, C. and Ackermann, F. (2004).
\newblock Cognitive mapping expert views for policy analysis in the public
  sector.
\newblock {\em European Journal of Operational Research}, 152(3):615--630.

\bibitem[Elmquist and Segrestin, 2009]{elmquist2009sustainable}
Elmquist, M. and Segrestin, B. (2009).
\newblock Sustainable development through innovative design: lessons from the
  {KCP} method experimented with an automotive firm.
\newblock {\em International Journal of Automotive technology and management},
  9(2):229--244.

\bibitem[Ferretti et~al., 2019]{ferretti2019studying}
Ferretti, V., Pluchinotta, I., and Tsouki{\`a}s, A. (2019).
\newblock Studying the generation of alternatives in public policy making
  processes.
\newblock {\em European Journal of Operational Research}, 273(1):353--363.

\bibitem[Friend and Hickling, 1987]{FriendHickling87}
Friend, J. and Hickling, A. (1987).
\newblock {\em Planning under pressure: the strategic choice approach}.
\newblock Pergamon Press, New York.

\bibitem[Frija et~al., 2016a]{frija2016farmers}
Frija, A., Chebil, A., and Speelman, S. (2016a).
\newblock Farmers' adaptation to groundwater shortage in the dry areas:
  improving appropriation or enhancing accommodation?
\newblock {\em Irrigation and Drainage}, 65(5):691--700.

\bibitem[Frija et~al., 2014]{frija2014critical}
Frija, A., Chebil, A., Speelman, S., and Faysse, N. (2014).
\newblock A critical assessment of groundwater governance in tunisia.
\newblock {\em Water Policy}, 16(2):358--373.

\bibitem[Frija et~al., 2016b]{frija2016gestion}
Frija, I., Frija, A., Marlet, S., Leghrissi, H., and Faysse, N. (2016b).
\newblock Gestion de l'usage d'une nappe par un groupement d'agriculteurs:
  l'exp{\'e}rience de {Bsissi Oued El Akarit} en {Tunisie}.
\newblock {\em Alternatives Rurales}, 4:on--line.

\bibitem[Graf et~al., 2007]{graf2007counselling}
Graf, W., Kramer, G., and Nicolescou, A. (2007).
\newblock Counselling and training for conflict transformation and
  peace-building: the transcend approach.
\newblock In Webel, C. and Galtung, J., editors, {\em Handbook of peace and
  conflict studies}, pages 123--142. Routledge, London.

\bibitem[Hardin, 1968]{hardin1974tragedy}
Hardin, G. (1968).
\newblock The tragedy of the commons.
\newblock {\em Science}, 162:1243--1248.

\bibitem[Hassenforder, 2023]{hassenforder2023accompagner}
Hassenforder, E. (2023).
\newblock Accompagner la gouvernance participative des eaux sousterraines pour
  une agriculture climatiquement intelligente en {Afrique du Nord}.
\newblock Technical report, BRIDGE C4S aquifer co-management activity.

\bibitem[Hassenforder et~al., 2024]{hassenforder2024participatory}
Hassenforder, E., Ferjani, A., and Trabelsi, F. (2024).
\newblock Participatory modeling of past, current and future groundwater
  governance: An experiment in {Aousja Ghar El Melh}, {Tunisia}.
\newblock {\em Futures}, 155:103281.

\bibitem[Hatchuel et~al., 2009]{hatchuel2009design}
Hatchuel, A., Le~Masson, P., and Weil, B. (2009).
\newblock Design theory and collective creativity: a theoretical framework to
  evaluate {KCP} process.
\newblock In {\em International conference on engineering design, ICED},
  volume~9, pages 24--27.

\bibitem[Hatchuel et~al., 2004]{hatchuel2004ck}
Hatchuel, A., Le~Masson, P., Weil, B., et~al. (2004).
\newblock {CK} theory in practice: lessons from industrial applications.
\newblock In {\em DS 32: Proceedings of DESIGN 2004, the 8th International
  Design Conference, Dubrovnik, Croatia}, pages 245--258.

\bibitem[Hatchuel and Weil, 2003]{hatchuel2003new}
Hatchuel, A. and Weil, B. (2003).
\newblock A new approach of innovative design: an introduction to {CK} theory.
\newblock In {\em DS 31: Proceedings of ICED 03, the 14th International
  Conference on Engineering Design, Stockholm}.

\bibitem[Hatchuel and Weil, 2009]{hatchuel2009ck}
Hatchuel, A. and Weil, B. (2009).
\newblock {CK} design theory: an advanced formulation.
\newblock {\em Research in engineering design}, 19:181--192.

\bibitem[Jacobi and Hobbs, 2007]{jacobi2007quantifying}
Jacobi, S. and Hobbs, B. (2007).
\newblock Quantifying and mitigating the splitting bias and other value
  tree-induced weighting biases.
\newblock {\em Decision Analysis}, 4(4):194--210.

\bibitem[Kazak{\c{c}}i et~al., 2009]{kazakcci2009formalization}
Kazak{\c{c}}i, O.~A. et~al. (2009).
\newblock A formalization of {CK} design theory based on intuitionist logic.
\newblock In {\em ICORD 09: Proceedings of the 2nd International Conference on
  Research into Design, Bangalore, India 07.-09.01. 2009}, pages 499--507.

\bibitem[Keeney, 1992]{Keeney92}
Keeney, R. (1992).
\newblock {\em Value-{F}ocused {T}hinking. {A} {P}ath to {C}reative {D}ecision
  {M}aking}.
\newblock Harvard {U}niversity {P}ress, Cambridge.

\bibitem[Keeney, 1994]{keeney1994creativity}
Keeney, R. (1994).
\newblock Creativity in decision making with value-focused thinking.
\newblock {\em Sloan Management Review}, 35:33--33.

\bibitem[Kelly, 1955]{kelly1955psychology}
Kelly, G. (1955).
\newblock {\em The psychology of personal constructs: A theory of personality}.
\newblock New York: Norton.

\bibitem[Mekki et~al., 2017]{mekki2017perceptions}
Mekki, I., Ghazouani, W., Closas, A., and Molle, F. (2017).
\newblock Perceptions of groundwater degradation and mitigation responses in
  the {Haouaria} region in {Tunisia}.
\newblock {\em Groundwater for Sustainable Development}, 5:101--110.

\bibitem[Mekki et~al., 2013]{mekki2013management}
Mekki, I., Jacob, F., Marlet, S., and Ghazouani, W. (2013).
\newblock Management of groundwater resources in relation to oasis
  sustainability: The case of the {Nefzawa} region in {Tunisia}.
\newblock {\em Journal of Environmental Management}, 121:142--151.

\bibitem[Mingers and Rosenhead, 2004]{mingers2004problem}
Mingers, J. and Rosenhead, J. (2004).
\newblock Problem structuring methods in action.
\newblock {\em European journal of operational research}, 152(3):530--554.

\bibitem[Minoia and Guglielmi, 2008]{minoia2008social}
Minoia, P. and Guglielmi, F. (2008).
\newblock Social conflict in water resource management and its environmental
  impacts in south-eastern {Tunisia}.
\newblock In Efe, R., Gravins, G., Öztürk, M., and Atalay, I., editors, {\em
  Natural environment and culture in the Mediterranean Region}, pages 257--270.
  Cambridge Scholars Publishing.

\bibitem[Molle and Closas, 2020]{molle2020comanagement}
Molle, F. and Closas, A. (2020).
\newblock Comanagement of groundwater: A review.
\newblock {\em Wiley Interdisciplinary Reviews: Water}, 7(1):e1394.

\bibitem[Ostrom, 1990]{ostrom1990governing}
Ostrom, E. (1990).
\newblock {\em Governing the commons: The evolution of institutions for
  collective action}.
\newblock Cambridge University Press.

\bibitem[Ostrom, 1999]{ostrom1999coping}
Ostrom, E. (1999).
\newblock Coping with tragedies of the commons.
\newblock {\em Annual review of political science}, 2(1):493--535.

\bibitem[Pluchinotta et~al., 2020]{pluchinotta2020integrating}
Pluchinotta, I., Giordano, R., Zikos, D., Krueger, T., and Tsoukias, A. (2020).
\newblock Integrating problem structuring methods and concept-knowledge theory
  for an advanced policy design: lessons from a case study in cyprus.
\newblock {\em Journal of Comparative Policy Analysis: Research and Practice},
  22(6):626--647.

\bibitem[Pluchinotta et~al., 2019]{pluchinotta2019design}
Pluchinotta, I., Kazak{\c{c}}i, A.~O., Giordano, R., and Tsouki{\`a}s, A.
  (2019).
\newblock Design theory for generating alternatives in public decision making
  processes.
\newblock {\em Group decision and negotiation}, 28:341--375.

\bibitem[P{\"o}yh{\"o}nen and H{\"a}m{\"a}l{\"a}inen, 1998]{poyhonen1998notes}
P{\"o}yh{\"o}nen, M. and H{\"a}m{\"a}l{\"a}inen, R. (1998).
\newblock Notes on the weighting biases in value trees.
\newblock {\em Journal of Behavioral Decision Making}, 11(2):139--150.

\bibitem[P{\"o}yh{\"o}nen et~al., 2001]{poyhonen2001behavioral}
P{\"o}yh{\"o}nen, M., Vrolijk, H., and H{\"a}m{\"a}l{\"a}inen, R. (2001).
\newblock Behavioral and procedural consequences of structural variation in
  value trees.
\newblock {\em European Journal of Operational Research}, 134(1):216--227.

\bibitem[Rosenhead, 1996]{rosenhead1996s}
Rosenhead, J. (1996).
\newblock What's the problem? an introduction to problem structuring methods.
\newblock {\em Interfaces}, 26(6):117--131.

\bibitem[Saidi et~al., 2013]{saidi2013groundwater}
Saidi, S., Bouri, S., and Dhia, H.~B. (2013).
\newblock Groundwater management based on gis techniques, chemical indicators
  and vulnerability to seawater intrusion modelling: application to the
  mahdia--ksour essaf aquifer, tunisia.
\newblock {\em Environmental earth sciences}, 70(4):1551--1568.

\bibitem[Simon, 1969]{simon1969sciences}
Simon, H.~A. (1969).
\newblock {\em The sciences of the artificial}.
\newblock MIT Press, Cambridge.

\bibitem[Simon, 1988]{simon1988science}
Simon, H.~A. (1988).
\newblock The science of design: Creating the artificial.
\newblock {\em Design Issues}, pages 67--82.

\bibitem[Soula et~al., 2023]{soula2023evaluation}
Soula, R., Chebil, A., Majdoub, R., Crespo, D., Albiac, J., and Kahil, T.
  (2023).
\newblock Evaluation of the impact of groundwater management policies under
  climate and economic changes in {Tunisia}.
\newblock {\em Water Economics \& Policy}, 9:2340005.

\bibitem[Soula et~al., 2021]{soula2021water}
Soula, R., Chebil, A., McCann, L., and Majdoub, R. (2021).
\newblock Water scarcity in the mahdia region of tunisia: Are improved water
  policies needed?
\newblock {\em Groundwater for Sustainable Development}, 12:100510.

\bibitem[Tosunlu et~al., 2023]{tosunlu2023conflict}
Tosunlu, B.~H., Guillaume, J.~H., and Tsouki{\`a}s, A. (2023).
\newblock Conflict transformation and management: from cognitive maps to value
  trees.
\newblock arxiv:2312.07838, arXiv.

\bibitem[Tringali et~al., 2017]{tringali2017insights}
Tringali, C., Re, V., Siciliano, G., Chkir, N., Tuci, C., and Zouari, K.
  (2017).
\newblock Insights and participatory actions driven by a socio-hydrogeological
  approach for groundwater management: the grombalia basin case study
  {Tunisia}.
\newblock {\em Hydrogeology Journal}, 25(5):1241.

\bibitem[Tsouki\`as, 2007]{Tsoukias07aor}
Tsouki\`as, A. (2007).
\newblock On the concept of decision aiding process.
\newblock {\em Annals of Operations Research}, 154:3 -- 27.

\bibitem[Tsouki\`as, 2008]{Tsoukias08ejor}
Tsouki\`as, A. (2008).
\newblock From decision theory to decision aiding methodology.
\newblock {\em European Journal of Operational Research}, 187:138 -- 161.

\bibitem[Vernoux and Horriche, 2019]{vernoux2019scenarios}
Vernoux, J.-F. and Horriche, F. (2019).
\newblock Scenarios of evolution of water consumption and groundwater
  management in the gabes jeffara--tunisia.
\newblock In {\em IC-SEWEN 2019}.

\bibitem[Von~Winterfeldt, 1987]{von1987value}
Von~Winterfeldt, D. (1987).
\newblock Value tree analysis: an introduction and an application to offshore
  oil drilling.
\newblock In Kleindorfer, P. and Kunreuther, H., editors, {\em Insuring and
  managing hazardous risks: from {Seveso} to {Bhopal} and beyond}, pages
  439--377. Springer, Berlin.

\end{thebibliography}


\end{document}